\documentstyle[ aps, epsfig]{revtex}

\draft

\begin{document}

\twocolumn[
\hsize\textwidth\columnwidth\hsize\csname@twocolumnfalse\endcsname

\title{Comment on 'Two Dimensional Ordering and Fluctuations in
$\alpha ^{\prime }$-NaV$_{2}$O$_{5}$'}
\author{S. Ravy$^{1}$ and J. E. Lorenzo$^{2}$}
\address{$^{1}$Laboratoire de physique des solides, CNRS UMR 8502, 
B\^{a}t. 510, Universit\'{e} Paris-sud, 91405 Orsay Cedex, France}
\address{$^{2}$European Synchrotron Radiation Facility, 38042 Grenoble Cedex, 
France}
\maketitle
\twocolumn]

\narrowtext

In a recent letter, B. G. Gaulin {\it et al.}\cite{Gaulin} present an x-ray
study of the $\alpha ^{\prime }$-NaV$_{2}$O$_{5}$ system. This compound,
widely considered as a 1/4-filled spin ladder, exhibits a phase transition
where charge and magnetic ordering occur (the ground state being
spin-singlet) in addition to a structural distortion with the modulation
wave vector (1/2,1/2,1/4). From their x-ray measurements above and below the
transition temperature T$_{c}$=33.1 K, the authors claim that the transition
is two-dimensional in nature and is due to a V$^{4+}$/V$^{5+}$-charge
ordering mechanism. We do not agree with these conclusions for three
reasons, as explained below.

The authors have measured the diffuse scattering above T$_{c}$ in the three
directions of the orthorhombic lattice (spin ladders are in the {\bf b }%
direction, regrouped in ({\bf a,b}){\bf \ }planes stacked in the {\bf c}
direction). In particular, scans in the {\bf c}$^{*}$ directions around the
superstructure spots, displayed in fig. 3 of \cite{Gaulin}, show the
presence of critical scattering due to fluctuations. Without any
quantitative analysis, the authors claim that these fluctuations develop
within $\sim $2 K above T$_{c}$, and that the scattering is very different
from that previously observed in \cite{Ravy}. The rod-like diffuse
scattering that is mentioned in \cite{Gaulin} has been also observed in \cite
{Ravy} and is due to the overlap of the broad Lorentzians located at $\Delta
l$=$\pm $0.5 from the measured one, and {\it not} due to a two-dimensional
regime of fluctuations, at least below $\sim $45 K. Indeed, from the
magnitude of the correlation lengths displayed in fig. 3 in \cite{Ravy},
which have been obtained by analysing consistently the data in {\it the
three directions}, one can see that they become very short-ranged along {\bf %
c}$^{*}$ $\sim $10 K above T$_{c}$. Therefore we do not see much difference
between the data reported in \cite{Gaulin} and \cite{Ravy}, and hence the
conclusion of \cite{Gaulin} regarding the critical scattering is doubtful,
at best. Especially, the statement that ''the three dimensional state is
built out of correlations developing between a-b planes which are already 
{\it well ordered}'' is based on a misleading data analysis.

Secondly, the authors claim that their results show ''two dimensional {\it %
charge ordering} to exist within the orthorhombic plane over an extensive
temperature range above T$_{c}$''. This assertion is incorrect in the sense
that x-ray mainly probe the lattice distortion and not the charge ordering
(which has a weak amplitude compared to the first one). An analogous
misconception is encountered in charge density wave (CDW) systems, in which
the satellite reflections appearing below the phase transition are not due
to the CDW itself, but to the associated lattice distortion. Evidencing
charge ordering by x-ray is a challenging task requiring {\it e.g}. the use
of anomalous scattering. This is not discussed in [1], where it is
implicitely assumed that the lattice mirrors the charge ordering.
Nevertheless, if the transition is triggered by a charge-ordering process
coupled to the lattice, it is reasonnable to assume that the anisotropy of
the lattice fluctuations observed by x-ray scattering is similar to that of
the charge-ordering. In this respect, the long range character of Coulomb
interactions involved in a charge-ordering process gives a natural
explanation of the three dimensional nature of the lattice fluctuations as
measured in \cite{Ravy}, contrary to what is stated in \cite{Gaulin}.

At last, the authors have measured the temperature dependence of the
superlattice peak below the transition temperature and obtained a $\beta $%
=0.17 critical exponent from a power law fit. From this value they conclude
''\ldots a very low value of $\beta $ strongly suggests ordering which is two
dimensional in nature''. The authors then provide an explanation borrowed
from results and theories of 2D magnetism (Refs. 20-24 of [1]), which is not
applicable to the case of {\it structural} phase transitions in solids. Such
low-dimensional {\it magnetic} phase transitions occur in 2D systems that
are actually embedded in a three-dimensional lattice but not coupled to the
lattice. In this respect, the comparison between the transition in $\alpha
^{\prime }$-NaV$_{2}$O$_{5}$ and the magnetic phase transition of 2D systems
(Refs. 20-24 of [1]), where a 2D regime of fluctuations is actually
observed, is not relevant. {\it Structural} phase transitions of
low-dimensional criticality have been observed in physically low-dimensional
compounds, as for instance freely suspended liquid-crystal films, but not in
real 3D solids. The proof of the 3D character of the transition is given by
the existence of three-dimensionnal critical fluctuations $\sim $10 K above T%
$_{c}$ \cite{Ravy}. The authors' claim on the identification of $\beta $%
=0.17 as a critical exponent of a 2D order parameter in NaV$_{2}$O$_{5}$ is
too adventurous.

In conclusion, we think that the analysis presented in Ref. \cite{Gaulin}
does not prove neither the two-dimensional character of the transition in NaV%
$_{2}$O$_{5}$, nor the charge ordering mechanism of the transition.

\end{document}